\documentclass[prd,preprintnumbers,showpacs,
showkeys,superscriptaddress,showpacs,nofootinbib,byrevtex,fleqn]{revtex4}
\usepackage{amsmath,amsfonts,amssymb,amscd,amsxtra,amsthm}
\usepackage{graphicx,wrapfig}
\usepackage{bm}

\begin{document}
\preprint{INHA-NTG-01/2012}
\title{Parity-violating $\pi N N$ coupling constant in the chiral
  quark-soliton model}

\author{Hee-Jung Lee}
\email{hjl@chungbuk.ac.kr}
\affiliation{Departament of Physics Education,
Chungbuk National University, Cheongju 361-763, Republic of Korea}

\author{Chang Ho Hyun}
\email{hch@daegu.ac.kr}
\affiliation{Department of Physics Education,
Daegu University, Gyeongsan 712-714, Republic of Korea}
\affiliation{School of Physics, Korea Institute for Advanced Study,
  Seoul 130-722, Republic of Korea} 

\author{Hyun-Chul Kim}
\email{hchkim@inha.ac.kr}
\affiliation{Department of Physics, Inha University,
Incheon 402-751, Republic of Korea}
\affiliation{Department of Physics, University of Connecticut,
  Storrs, CT 06269, U.S.A.} 
\affiliation{School of Physics, Korea Institute for Advanced Study,
  Seoul 130-722, Republic of Korea}

\date{June 2012}

\begin{abstract}
We investigate the parity-violating $\pi NN$ Yukawa coupling constant
$h^1_{\pi NN}$ within the framework of the SU(2) chiral quark-soliton model,
based on the $\Delta S=0$ effective weak Lagrangian derived within the
same framework. We find that the parity-violating $\pi NN$ coupling
constant is about $1\times10^{-8}$ at the scale of 1 GeV. The results
of $h_{\pi NN}^1$ turn out to be sensitive to the Wilson coefficient. 
We discuss how the gluonic renormalization supresses the
parity-violating $\pi NN$ coupling constant.
\end{abstract}
\maketitle

\section{Introduction \label{sec:intro}}
The parity-violating (PV) hadronic processes in low-energy regions
have been one of the most fundamental issues in nuclear and hadronic
physics for long time (see a recent review~\cite{Holstein:2009zz} for
some historical and phenomenological background). However, the weak
interactions of hadrons are yet poorly understood because of the
strong interaction, compared to lepton-lepton or lepton-hadron weak
processes. For example, the long-standing puzzle of the $\Delta I =
1/2$ rule in strangeness-changing weak interactions indicates that the
effect of the strong interaction in weak processes raises a
non-trivial
problem~\cite{Kambor:1989tz,Buchalla:2008jp,Lellouch:2011qw}. It is 
even more difficult to study parity-violating nuclear processes
because of experimental feasibility and theoretical complication
caused by the nonperturbative strong interaction of quarks and
gluons. The standard model (SM) asserts that charged weak boson
exchange induces flavor-changing weak interactions whereas the neutral
current conserves the flavor. The basic ingredient to describe
low-energy hadronic weak processes is the quark current-current
interaction with $W$ and $Z$ bosons. However, in order to describe
low-energy phenomena below 1 GeV, one has to scale down
this interaction from the mass scale of the $W$ and $Z$. In
the course of this scaling, the quark-gluon interactions are encoded
in the Wilson coefficients by the renormalization group
equation~\cite{Gaillard:1974nj,Shifman:1975tn,Buchalla:1995vs,
Desplanques:1979hn}, which, however, explains only a perturbative part
of the strong interaction.

Desplanques, Donoghue, and Holstein (DDH)~\cite{Desplanques:1979hn}
suggested that hadronic and nuclear PV processes can be described by
one-boson exchange such as $\pi$-, $\rho$-, and
$\omega$-exchanges~\cite{Desplanques:1979hn,Miller:1982ij,Dubovik:1986pj}
\`{a} la the strong nucleon-nucleon ($NN$) potential.
The main factors of the PV $NN$ potential are the 
seven weak meson-$NN$ coupling constants, i.e. $h_{\pi NN}^1$,
$h_{\rho NN}^0$, $h_{\rho NN}^1$, $h_{\rho NN}^2$, $h_{\omega NN}^0$,
$h_{\omega  NN}^1$, and $h_{\rho NN}^{\prime 1}$,
where superscripts denote the isospin difference $\Delta I$.
Among these coupling constants, it is of utmost importance to
understand the PV $\pi NN$ coupling constant, because it governs the
long-range part of the PV $NN$ interaction, so that it plays the most
signicant role in explaining the PV nuclear processes.
The PV $\pi NN$ coupling constant can in principle be extracted from
various PV reactions $\bm n p\to
d\gamma$~\cite{Cavaignac:1977uk,Snow:2000az,Gericke:2011zz}, and
$^{18}\mathrm{F}^*\to ^{18}
\mathrm{F}$~\cite{Barnes:1978sq,Bizzeti:1980ru,Page:1987ak} but is
fraught with large undertainties. We refer to a recent
review~\cite{Haxton:2008ci} for the present status of hadronic PV
experiments. It has been also calculated in various theoretical
frameworks: the $\mathrm{SU(6)}_W$ quark
model~\cite{Desplanques:1979hn,Feldman:1991tj} with the 
effective weak Hamiltonian, the Skyrme model~\cite{Kaiser:1989fd, 
  Kaiser:1989ah,Meissner:1998pu}, and QCD sum
rules~\cite{Henley:1995ad}, and so on. Even though a great amount of
efforts was made on understanding $h^1_{\pi NN}$ experimentally as well as
theoretically, its quantitative value is still elusive. 

In the present work, we investigate the PV $\pi NN$ coupling
constant, $h^1_{\pi NN}$, within the framework of the SU(2) chiral
quark-soliton model ($\chi$QSM) which is an effective chiral
model for QCD in the low-energy region with constituent quarks and the 
pseudoscalar mesons as the relevant degrees of freedom. The model
respects the spontaneous breakdown of chiral symmetry and describes
baryons fully relativistically. Moreover, it is deeply related to the
QCD vacuum based on instantons~\cite{CQSM:Diakonovlecture} and
contains only a few free parameters. These parameters can mostly be
fixed to the meson masses and meson decay constants in the mesonic
sector. The only remaining free parameter is the constituent quark
mass or dynamical quark mass that is also fixed by reproducing the
electric properties of the proton. The $\chi$QSM was successful in
describing lowest-lying baryon
properties~\cite{Christov:1995vm}. Furthermore, the renormalization
scale for the $\chi$QSM is naturally given by the cut-off parameter
for the regularization which is about $0.36\,\mathrm{GeV^2}$. Note
that it is implicitly related to the inverse of the size of instantons 
($\overline{\rho}\approx 0.35\,\mathrm{fm}$)
~\cite{Diakonov:1983hh,Diakonov:1985eg}.
This renormalization scale is very important in general, 
because the essential feature of the PV hadronic interactions comes
from the effective weak Hamiltonian that has a specific scale
dependence, as mentioned previously. Thus, the matching of this scale
consists of an essential part in investigatng any nonleptonic
decays and PV hadronic processes. 

While the $\chi$QSM provides a plausible framework to study the PV 
$\pi NN$ coupling constant, there are at least two theoretical
difficulties. Firstly, the effective weak Hamiltonian has two-body
operators and one has to treat the four-point correlation
functions in order to compute the PV $\pi NN$ coupling 
constant. Secondly, since the momentum-dependent dynamical quark 
mass is known to play a significant role in describing $K\to \pi\pi$ 
nonleptonic decays~\cite{Franz:1999ik,Franz:1999wr}, one can expect
that it would also contribute to $h^1_{\pi NN}$ substantially. This is in
particular important, because a certain amount of non-perturbative
effects is reflected in the momentum-dependent quark mass, which
arises from the zero mode of instantons. However, it
is very difficult to handle these problems 
in the self-consistent $\chi$QSM. In order to circumvent all
technical difficulties in the self-consistent approach, we will use
the gradient expansion to calculate $h^1_{\pi NN}$, taking the limit of a
large soliton size, so that valence quarks in a nucleon plunge into
the Dirac sea and the soliton emerges as a topological
one~\cite{Christov:1995vm}, which is quite similar to a
skyrmion. Equivalently, we can start directly from the $\Delta S=0$
effective weak chiral Lagrangian derived in Ref.~\cite{Lee:2004tr} and
quantize the chiral soliton collectively. Then, we introduce a
physical pion through quantum fluctuations around the soliton
field. This procedure will lead to the results for $h^1_{\pi NN}$ without
fitting any parameter. In the present work, we will restrict ourselves
the SU(2) case for simplicity and will concentrate on how the
low-energy constants (LECs) found in Ref.~\cite{Lee:2004tr}  feature
the PV $\pi NN$ coupling constant. 

This paper is organized as follows:
In Section \ref{sec:lag}, we describe briefly a general formalism for
the derivation of the PV weak $\pi NN$ coupling constant.
In Section \ref{sec:res} we present the numerical results for
$h^1_{\pi NN}$ and discuss the role of the LECs of the
$\Delta S=0$ effective weak chiral Lagrangian. The last section is
devoted to the summary and outlook of this work. 

\section{General formalism \label{sec:lag}}
In this Section, we will show how to incorporate the $\Delta S=0$
effective weak Hamiltonian into the effective chiral action. We employ
the $\Delta S=0$ effective weak Hamiltonian derived in
Ref.~\cite{Desplanques:1979hn}.  The Hamiltonian reads
\begin{eqnarray}
{\cal H}_{W}^{\Delta S=0} &=&
\frac{G_F}{\sqrt{2}}\cos\theta_c\sin\theta_c \left[
\sum_{i=1}^2(\alpha_{ii}{\cal O}(A_i^\dagger, A_i)
+\beta_{ii}{\cal O}(A_i^\dagger t_A, A_i t_A)+{\rm
  h.c.}))\right. \nonumber\\ 
&& \left. +\, \sum_{i,j=1}^2(\gamma_{ij}{\cal O}(B_i^\dagger, B_j)
+\rho_{ij}{\cal O}(B_i^\dagger t_A, B_j t_A)) \right],
\label{Eq:Heff}
\end{eqnarray}
where the operator $\mathcal{O}(M_i,N_i)$ is defined as a two-body operator
$\mathcal{O}(M_i,N_i) \equiv -{\psi}^\dagger\gamma_\mu\gamma_5 M_i \psi
{\psi}^\dagger\gamma^\mu N_i\psi$ in Euclidean space, and $t_A$ denotes
the generator of the color $SU(3)$ group, normalized as ${\rm tr}\,
t_At_B=2\delta_{AB}$.  The definitions of the matrices $A_i$ and
$B_i$, and the coefficients $\alpha$, $\beta$, $\gamma$ and $\rho$ can
be found in \cite{Desplanques:1979hn}. These coefficients are the
functions of the scale-dependent Wilson coefficient $K(\mu)$ defined
as
\begin{equation}
K(\mu) \equiv
\bigg(1+\frac{g^2(\mu^2)}{16\pi^2}b\ln\frac{M_W^2}{\mu^2}\bigg),
\end{equation}
where $g(\mu^2)$ denotes the strong running coupling constant, $\mu$
stands for the renormalization point that specifies the energy scale,
$b=11-2 N_f/3$, and $M_W$ is the mass of the $W$ boson.  The
coefficient $K$ encodes the effect of the strong interaction from
perturbative gluon exchanges.

The four-quark operators are expressed generically by
\begin{equation}
{\cal Q}^i(x)=-{\psi}^\dagger(x)\Gamma_1^i\psi(x)
{\psi}^\dagger(x)\Gamma_2^i\psi(x)\ ,
\end{equation}
where $i(=1,\cdots, 12)$ labels each four-quark operator in the
effective weak Hamiltonian and $\Gamma_{1(2)}^i$ consist of the
Dirac gamma and flavor matrices.  Thus, the effective weak 
Hamiltonian can be rewritten as follows:
\begin{equation}
{\cal H}^{\Delta S=0}_{W}
=\sum_{i=1}^{12}{\cal C}_i {\cal Q}^i(x),
\end{equation}
where ${\cal C}_i$ denotes $\alpha$, $\beta$, $\gamma$ and
$\rho$ according to Eq.~(\ref{Eq:Heff}).

In order to derive $h_{\pi NN}^1$ in the $\chi$QSM, we have to solve
the following matrix element: 
\begin{equation}
 \left\langle N \left| {\cal H}_{W}^{\Delta S=0} \right| \pi^a
  N\right\rangle \;=\; \sum_{i=1}^{12}{\cal C}_i \left \langle N \left
  |{\cal Q}^i(z) \right| \pi^a N\right\rangle \;=\;
\sum_{i=1}^{12}{\cal C}_i \int d^4 \xi e^{ik\cdot \xi} (k^2+m_\pi^2)
\left \langle N \left |{\cal T} [{\cal Q}^i(z)\pi^a(\xi) ]
  \right| N\right\rangle.
\label{eq:mat1} 
\end{equation}
The nucleon state is defined in terms of the Ioffe-type current in
Euclidean space ($x_{0}=-ix_{4}$):
\begin{equation}
|N(p_{1})\rangle \;=\; \lim _{y_{4}\rightarrow -\infty }
e^{p_{4}y_{4}}{\mathcal{N}}^{*}(p_{1})\int d^{3}y
e^{i{\bm p}_{1}\cdot {\bm y}}J^{\dagger }_{N}(y)|0\rangle,\;\;\;
\langle N(p_{2})| \;=\; \lim _{x_{4}\rightarrow +\infty }
e^{-p_{0}x_{4}}{\mathcal{N}}(p_{2})\int d^{3}x
e^{-i{\bm p}_{2}\cdot {\bm x}}\langle 0| J_{N}(x).  
\end{equation}
The nucleon current \( J^{\dagger }_{N}\,(J_{N}) \) plays a role of creating
(annihilating) nucleons. The ${\mathcal{N}}^{*}$ (${\mathcal{N}}$)
represensts the normalizing factor depending on the initial (final)
momentum. The  $J^{\dagger }_{N}$ ($J_{N}$) consists of $N_{c}$
quarks: 
\begin{equation}
J_{N}(x)\; =\; \frac{1}{N_{c}!}\epsilon ^{c_{1}c_{2}\cdots c_{N_{c}}}
\Gamma ^{s_{1} s_{2}\cdots
s_{N_{c}}}_{(TT_{3}Y)(JJ_{3}Y_{R})}\psi _{s_{1}c_{1}}(x)
\cdots \psi _{s_{N_{c}}c_{N_{c}}}(x)\,,
\end{equation}
where $s _{1}\cdots s _{N_{c}}$ and $c_{1}\cdots c_{N_{c}}$
denote respectively spin-isospin and color indices. 
The $\Gamma ^{\{s \}}_{(TT_{3}Y)(JJ_{3}Y_{R})}$ are matrices with
the quantum numbers \( (TT_{3}Y)(JJ_{3}Y_{R}) \). For the 
nucleon, \( T=1/2 \), \( Y=1 \) and \( J=1/2 \). The right hypercharge will
be constrained by the baryon number. The creation baryon current is
written as
\begin{equation}
J^{\dagger }_{N}(y)\; =\; \frac{1}{N_{c}!}\epsilon ^{c_{1}c_{2}
\cdots c_{N_{c}}}\Gamma ^{s_{1}s_{2}\cdots s_{N_{c}}*
}_{(TT_{3}Y)(JJ_{3}Y_{R})}\left( -i\psi ^{\dagger } 
\gamma _{4}\right) _{s _{N_{c}}c_{N_{c}}}(x)\cdots
\left( -i\psi ^{\dagger }\gamma _{4}\right) _{s _{1}c_{1}}(x).
\end{equation}
The partial conservation of the axial-vector current (PCAC) being
considered, the matrix elements in Eq.(\ref{eq:mat1}) can be related
to the following four-point correlation function
\begin{equation}
\label{Eq:correl}
\lim_{y_{0}\rightarrow -\infty \atop x_{0}\rightarrow +\infty }
\sum_{i=1}^{12}{\cal C}_i \langle 0 | {\mathcal T}[J_{N}(x){\cal
  Q}^i(z) \partial_\mu A_\mu^a (\xi) J^{\dagger }_{N}(y) ]|0\rangle\; =\;
\lim _{y_{0}\rightarrow -\infty
\atop x_{0}\rightarrow +\infty }{\mathcal{K}},
\end{equation}
where $A_\mu^a$ stands for the axial-vector current. In the $\chi$QSM,
the correlation function $\mathcal{K}$ can be expressed as a
functional integral 
\begin{equation}
\mathcal{K} \;=\; \frac1{\mathcal{Z}} \int D\psi D\psi^\dagger DU
J_{N}(x){\cal Q}^i(z) \partial_\mu A_\mu^a (\xi) J^{\dagger }_{N}(y)
\exp\left[\int d^4x \psi^\dagger \left(i\rlap{/}{\partial} + i
    \sqrt{M(-\partial^2)}U^{\gamma_5}\sqrt{M(-\partial)^2}\right)
  \psi\right]
\label{eq:corr2}
\end{equation}
in the chiral limit, where $M(-\partial^2 )$ denotes the
momentum-dependent dynamical quark mass and $U^{\gamma_5}$ represents
the chiral field defined as
\begin{equation}
U^{\gamma_5} \;=\; \frac{1+\gamma_5}{2} U + \frac{1-\gamma_5}{2}
U^\dagger
\end{equation}
with the Goldstone boson field $U=\exp(i\lambda^a \pi^a/f_\pi)$.

It is, however, extremely complicated to solve Eq.~(\ref{eq:corr2})
numerically, since the PV $\pi NN$ coupling constant involves the
two-body quark operators $\mathcal{Q}^i$ and the axial-vector one,
which will lead to laborious triple sums in quark levels already at
the leading order. Moreover, the momentum-dependent quark mass, which
is known to be of great significance in describing nonleptonic 
processes~\cite{Franz:1999ik}, introduces in addition technical
difficulties~\cite{Broniowski:2001cx}.  
One way to avoid these complexities is to use a gradient expansion
taking $(\rlap{/}{\partial}U/M)\ll 1$~\cite{Diakonov:1987ty}  
or equivalently is to start from the effective weak chiral Lagrangian
already derived in Ref.~\cite{Lee:2004tr}. Note that 
though we did not carry out the derivative expansion to order $p^4$,
it is not difficult to estimate how large the corresponding LECs could 
be. In Ref.~\cite{Franz:1999wr}, the $\Delta S=1$ effective weak
chiral Lagrangian to order $p^4$ was investigated in the case of the
local chiral quark model. As one can see, all of the LECs are
order-of-magnitude smaller than the $\mathcal{O}(p^2)$ 
LECs. In this sense, even though we go further beyond the leading
order, the contribution from higher derivative terms will not enhance
or suppress $h_{\pi NN}$ much. It will be at most below
$(5-10)\,\%$. Thus, we will use the $\Delta S=0$ effective weak chiral
Lagrangian derived in Ref.~\cite{Lee:2004tr} as our starting point,
instead of dealing with Eq.~(\ref{eq:corr2}).
Nevertheless, the present approach goes beyond the previous analyses
in the Skyrme
model~\cite{Kaiser:1989fd,Kaiser:1989ah,Meissner:1998pu}, 
because the present scheme incorporates properly 
the effects of the perturbative quark-gluon strong interaction in the 
derivation of the PV $\pi NN$ coupling constant.  

The leading-order (LO) term of the $\Delta S=0$ effective weak chiral
Lagrangian in the large $N_c$ can be expressed in terms of the vector
and axial-vector currents 
\begin{eqnarray}
{\cal L}_{\rm LO}
&=& 2\bigg(
\tilde{\alpha}_{11}\sum_{i=1}^2V_\mu^i A^{i \mu}
+\tilde{\alpha}_{22}\sum_{i=4}^5V_\mu^i A^{i \mu}\bigg)
+\bigg [9\tilde{\gamma}_{11} V_\mu^0 A^{0 \mu}
+3\tilde{\gamma}_{12}
\bigg(-V_\mu^0+2V_\mu^3+\frac{2}{\sqrt{3}}V_\mu^8\bigg)
A^{0 \mu}
\cr
&+& 3\tilde{\gamma}_{21}V^{0 \mu}
\bigg(-A_\mu^0+2A_\mu^3+\frac{2}{\sqrt{3}}A_\mu^8\bigg)
+\tilde{\gamma}_{22}
\bigg(-V_\mu^0+2V_\mu^3+\frac{2}{\sqrt{3}}V_\mu^8\bigg)
\bigg(-A^{0 \mu}+2A^{3 \mu} +\frac{2}{\sqrt{3}}A^{8 \mu} \bigg)
\bigg],
\label{eq:L-leading}
\end{eqnarray}
where the vector and axial-vector currents are defined as
\begin{equation}
V_\mu^a=\frac{f_\pi^2}{2}{\rm Tr}[T^a(R_\mu+L_\mu)]\,, \;\;\;\;
A_\mu^a=\frac{f_\pi^2}{2}{\rm Tr}[T^a(R_\mu-L_\mu)]
\end{equation}
in terms of $L_\mu=iU^\dagger \partial_\mu U$, $R_\mu=iU \partial_\mu
U^\dagger$, and $T^a = \left( \frac{1}{3},\, \frac{\lambda^1}{2},\,
\cdots,\, \frac{\lambda^8}{2} \right)$.
The parameter $f_\pi$ stands for the pion decay
constant $f_\pi=93$ MeV.
The explicit expressions for the coefficients $\tilde{a}_{ij}$ ($a =
\alpha,\, \beta,\, \gamma,\, \rho$) can be found in
Ref.~\cite{Lee:2004tr}.

The classical soliton field $U_0$ is assumed to have a structure of
the trivial embedding of the SU(2) hedgehog field as
\begin{eqnarray}
U_0=\left(\begin{array}{cc}\exp(i{\bm\tau}\cdot\hat{\bm r}P(r))& 0 \\
0&1 \end{array}\right)
\label{eq:embed}
\end{eqnarray}
with the profile function of the soliton $P(r)$. This classical
soliton field can be fluctuated in such a way that the pion field can
be coupled to a weak two-body operator
\begin{eqnarray}
U&=&\exp(i{\bm \tau}\cdot{\bm \pi}/2)\, U_0\, \exp(i{\bm
  \tau}\cdot{\bm \pi}/2).
\end{eqnarray}
Similarly, the vector and the axial-vector currents transform as
\begin{equation}
A_\mu^a \;=\;
\tilde{A}_{\mu}^a+\frac{1}{f_\pi} f^{abi} \tilde{V}^b_{\mu} \pi^i, \;\;\;\;\;
V_\mu^a \;=\;
\tilde{V}_{\mu}^a+\frac{1}{f_\pi} f^{abi} \tilde{A}^b_{\mu} \pi^i,  
\end{equation}
where the indices $a,\, b=1,\, \cdots ,\, 8$ and $i =$ 1, 2, 3.
The current with a tilde indicates that arising from the background
soliton field.

Since the PV $\pi NN$ interaction Lagrangian is expressed as
\begin{equation}
{\cal L}^\pi_{\rm pv}= - \frac{1}{\sqrt{2}}h_{\pi NN}^1\bar{\Psi}_N
(\bm{\tau}\times\bm{\pi})_3\Psi_N\,,
\label{eq:pvpnn}
\end{equation}
which is linear in the pion field and defined in the SU(2) flavor
space (proportional to $(\bm{\tau}\times\bm{\pi})_3$),
one can easily see that the term $\sum_{i=4}^5 V^i_\mu A^i_\mu$ does
not contribute to the PV $\pi NN$ Lagrangian. Moreover, since $f^{8bi}
= f^{0bi} = 0$, the pion fields for the PV $\pi NN$ Lagrangian
can survive in the vector
and axial-vector currents only when $a = i$. Writing them
explicitly, we have
\begin{equation}
A_\mu^i \;=\; \tilde{A}_{\mu}^i+\frac{1}{f_\pi}({\tilde{\bm V}}_{\mu}
\times\bm{\pi})^i, \;\;\;\;\;
V_\mu^i \;=\;
\tilde{V}_{\mu}^i+\frac{1}{f_\pi}({\tilde{\bm A}}_{\mu}
\times\bm{\pi})^i.  
\end{equation}
Considering the terms contributing to the PV $\pi NN$ vertex, we
obtain for the LO Lagrangian
\begin{eqnarray}
{\cal L}^\pi_{\rm LO} &=&
\tilde{\alpha}_{11}\, \sum_{i=1}^2 V^i_\mu\, A^{i \mu}
+ (3 \tilde{\gamma}_{12}-\tilde{\gamma}_{22})\, V_{\mu}^3\, A^{0\mu}
+ (3 \tilde{\gamma}_{21}-\tilde{\gamma}_{22})\, V^0_\mu\, A^{3 \mu}
\nonumber\\
&&+ 2 \tilde{\gamma}_{22}\, V^3_\mu\, A^{3 \mu}
+\frac{2}{\sqrt{3}} \tilde{\gamma}_{22}\,
(V^3_\mu\, A^{8 \mu} + V^8_\mu\, A^{3 \mu})+\, (V\leftrightarrow A).
\label{eq:pilag1}
\end{eqnarray}
Extracting the terms linear in the pion field from
Eq.~(\ref{eq:pilag1}) and rearranging them,
we finally derive the LO PV $\pi NN$ Lagrangian:
\begin{eqnarray}
{\cal L}^\pi_{\rm LO} &=& \frac{1}{f_\pi} \left\{
(- \tilde{\alpha}_{11} + 2 \tilde{\gamma}_{22} )
\left[V^3_{\mu} (\bm{V}^{\mu} \times \bm{\pi})^3
+ A^3_{\mu}(\bm{A}^{\mu} \times \bm{\pi})^3
\right] \right. \nonumber \\
& & + (3 \tilde{\gamma}_{21} - \tilde{\gamma}_{22})
V^0_\mu (\bm{V}^{\mu} \times \bm{\pi})^3
+ (3\tilde{\gamma}_{12} - \tilde{\gamma}_{22})
A^0_\mu (\bm{A}^{\mu} \times \bm{\pi})^3
\nonumber \\ & & \left.
+ \frac{2}{\sqrt{3}} \tilde{\gamma}_{22} \left[
A^8_\mu (\bm{A}^{\mu} \times \bm{\pi})^3
+ V^8_\mu (\bm{V}^{\mu} \times \bm{\pi})^3
\right] \right\}+\, \bigg({\bm O}^{0,3,8}\leftrightarrow ({\bm
O}\times {\bm \pi})^3\bigg). 
\label{eq:pilag2}
\end{eqnarray}
For simplicity, we have omitted the tildes in the currents.

In a similar manner, the next-to-leading order (NLO) effective weak chiral
Lagrangian in the large $N_c$ expansion derived in \cite{Lee:2004tr}
yields the Lagrangian for the PV $\pi NN$ vertex as
\begin{eqnarray}
{\cal L}_{\rm NLO}^\pi &=&\frac{1}{N_c \, f_\pi} \bigg\{
- (\tilde{\alpha}_{11}+2\tilde{\beta}_{11})
(\bm{\Lambda}_3\times\bm{\pi})_3
+ (\tilde{\alpha}_{22}+2\tilde{\beta}_{22})
\bigg[(\bm{\Lambda}_4\times\bm{\pi})_4
+(\bm{\Lambda}_5\times\bm{\pi})_5\bigg] \cr
&& +\, 3\left(\frac{4{\cal I}_1{\cal I}_3}{{\cal I}^2_2} + 1\right)
(\tilde{\gamma}_{12}+2\tilde{\rho}_{12})
(\bm{\Lambda}_0\times\bm{\pi})_3
+ 3\left(\frac{4{\cal I}_1{\cal I}_3}{{\cal I}^2_2} - 1\right)
(\tilde{\gamma}_{21}+2\tilde{\rho}_{21})
(\bm{\Lambda}_0\times\bm{\pi})_3 \cr
&& +\, 2(\tilde{\gamma}_{22}+2\tilde{\rho}_{22})
\bigg[(\bm{\Lambda}_3\times\bm{\pi})_3
+\frac{1}{\sqrt{3}}(\bm{\Lambda}_8\times\bm{\pi})_3\bigg]\
\bigg\}\,,
\label{eq:NLO}
\end{eqnarray}
where
\begin{eqnarray}
\bm{\Lambda}_a &\equiv&
\frac{f^4_\pi}{4} \mbox{Tr}\left[(R_\mu\lambda_a R^\mu+L_\mu\lambda_a L^\mu)
\bm{\tau}\right] \ \ \ {\rm for}\ \ a=0,\, 3,\, 8, \\
(\bm{\Lambda}_a\times\bm{\pi})_a &\equiv&
\frac{f^4_\pi}{4} \mbox{Tr}\left[(R_\mu\lambda_a R^\mu+L_\mu\lambda_a L^\mu)
f_{abi}\lambda_b\pi_i\right] \ \ \ {\rm for}\ \ a=4,\ 5 ,
\end{eqnarray}
and $\lambda_0$ is defined as the unit matrix in SU(3) divided by 3.  
The integrals $\mathcal{I}_i$ in Eq.(\ref{eq:NLO}) were already 
evaluated in Ref.~\cite{Lee:2004tr} and are expressed as 
\begin{eqnarray}
{\cal I}_{1}&=&-\int \frac{d^4 k}{(2\pi)^4}\ \frac{M(k)}{k^2+M^2(k)}
=\frac{\left\langle\overline{\psi}{\psi}\right\rangle_ M}{4N_c},
\label{eq:coefficient1} \\
{\cal I}_{2}&=&\int \frac{d^4 k}{(2\pi)^4}\ \frac{M^2(k)
-\frac{k^2}{2}M(k)\tilde{M}^\prime}{(k^2+M^2(k))^2} =
\frac{f_\pi^2}{4N_c}, \label{eq:coefficient2}\\
{\cal I}_{3}&=&\int \frac{d^4 k}{(2\pi)^4}\
\bigg[\frac{\frac{1}{4}\tilde{M}^{\prime\prime}k^2
+\frac{1}{2}\tilde{M}^\prime-\frac{\tilde{M}^{\prime2}}{8M}k^2}{k^2+M^2(k)}
-\frac{M+M^2\tilde{M}^\prime+\frac{k^2}{2}M^2\tilde{M}^{\prime\prime}
+\frac{1}{2}k^2M\tilde{M}^{\prime2}
+\frac{k^2}{4}\tilde{M}^\prime}{(k^2+M^2(k))^2} \cr
&&
\hspace{1.5cm} +\,k^2\frac{\frac{1}{2}M+2M^2\tilde{M}^\prime
+M^3\tilde{M}^{\prime2}}{(k^2+M^2(k))^3}\bigg],
\label{eq:coefficient3}
\end{eqnarray}
where $\langle \bar{\psi}\psi\rangle_M$ denotes the quark condensate in
Minkowski space and $\tilde{M}'=(dM(k)/dk)/2k$.

The next step is to carry out the zero-mode collective quantization of
the soliton
\begin{equation}
U_0(\vec{x})\rightarrow U(\vec{x},t)=R(t)U_0(\vec{x})R^\dagger(t)\ ,
\end{equation}
where $R(t)$ stands for the unitary time-dependent SU(3) orientation
matrix of the soliton $R(t)=\exp(i\Omega^a(t)\lambda^a/2)$ with its
angular velocity $\Omega^a(t)$ that is of order
$\mathcal{O}(1/N_c)$. Each current is transformed as 
\begin{eqnarray}
V^a_0&=&\frac{f_\pi^2}{2}{\rm Tr}\bigg(i\frac{\lambda^a}{2}R
[[U_0, R^\dagger \dot{R}], U_0^\dagger]R^\dagger\bigg)\cr
&=& \frac{f_\pi^2}{2}[1-\cos P(r)]D^{a\alpha}\Omega^{\alpha}
+f_\pi^2\sin^2 P(r)D^{ai} \Omega^i
-f_\pi^2\sin^2P(r)(\hat{\bm r}\cdot{\bm \Omega})D^{ai}\hat{r}^i\ ,
\\
V^a_i&=&\frac{f_\pi^2}{2}{\rm Tr}\bigg(i\frac{\lambda^a}{2}R
[U_0,\partial_iU_0^\dagger]R^\dagger\bigg)
=f_\pi^2\frac{\sin^2P(r)}{r}f_{ijk}\hat{r}_jD^{ak}\ ,
\\
A^a_0&=&-\frac{f_\pi^2}{2}{\rm Tr}\bigg(i\frac{\lambda^a}{2}R
\{[U_0, R^\dagger \dot{A}], U_0^\dagger\} R^\dagger\bigg)
\nonumber\\
&=&f_\pi^2\bigg(\sin P(r)\cos P(r)\hat{r}^i\epsilon_{ijk}D^{ak}\Omega^j
+\sin P(r)\hat{r}^if_{i\alpha\beta} D^{a\beta}\Omega^\alpha\bigg)\ ,
\\
A^a_i&=&\frac{f_\pi^2}{2}{\rm Tr}\bigg(i\frac{\lambda^a}{2} R
\{U_0,\partial_iU_0^\dagger\} R^\dagger\bigg)
=f_\pi^2\bigg[\frac{\sin 2P(r)}{2r}\delta_{ij}
+\bigg((P^{\prime})^2-\frac{\sin 2P(r)}{2r}\bigg)
\hat{r}_i\hat{r}_j\bigg]D^{aj}\ ,
\end{eqnarray}
where Italic~(Greek) indices run over $1,\,2,\,3$ ($4,\cdots,7$),
respectively, and dot~(prime) means the derivative with respect to
time (radius), respectively. The Wigner $D$ functions and the angular
velocity are defined as
\begin{equation}
D^{ab}(R) \;=\;
\frac{1}{2}{\rm Tr}(\lambda^a R\lambda^b
R^\dagger),\;\;\;\;\;
R^\dagger \dot{R} \;=\; \frac{i}{2}\lambda^a\Omega^a.  
\end{equation}

Before we proceed the calculation of $h_{\pi NN}^1$, we want to
emphasize that we will investigate $h_{\pi NN}^1$ first in the SU(2)
case in this work. Of course, the strange quarks may still
play a certain role in describing $h_{\pi NN}^1$. In fact,
Ref.~\cite{Kaplan:1992vj} showed that the strange quark operator
$(\bar{q}\lambda^3 \gamma_\mu q)(\bar{s}\gamma_\mu\gamma_5 s )$
induced by $Z^0$ exchange could     
contribute significantly to the $NN$ coupling constant. The main 
argument of Ref.~\cite{Kaplan:1992vj} lies in the fact that the
$\Delta I=1$ operator proportional to $h_{\pi NN}^1$ can be related to
the $\Delta S=1$ operator by an SU(3) rotation followed by an isospin
rotation. Then, it was found that the linear combination of the
strange operators made a large contribution to $h_{\pi NN}^1$, which
indicates that it has large matrix elements in the nucleon state.
The SU(3) Skyrme model came to the similar conclusion that the four
quark operators with the strange quark contributed to $h_{\pi NN}^1$ 
significantly~\cite{Meissner:1998pu} because of the induced kaon
field. Note, however, that Ref.~\cite{Meissner:1998pu} has not 
used the renormalized effective weak Hamiltonian but started from the 
bare Hamiltonian. On the other hand, in a recent lattice
study~\cite{Wasem:2011tp}, the strange quark operators can only
contribute to the quark-loop diagrams for which the signal-to-noise
ratio remains far too small to bring out any reasonable signal, so
that they were neglected. Moreover, recent findings have it that the
content of strange quarks in the nucleon in the vector channel is
negligible small~\cite{Ahmed:2011vp} and that the strangeness in the
scalar and  axial-vector channels is still hampered by 
uncertainties~\cite{Ellis:2008hf}. Thus, it is still too early to
reach a conclusion on the contribution of strange quark operators to
$h_{\pi NN}^1$. In the present work, we will concentrate on the case of
SU(2), since it does not vanish even in SU(2). As we will discuss
later in detail, this finite result is distinguished from that of the
SU(2) Skyrme model~\cite{Shmatikov:1989mf} in which $h_{\pi NN}^1$
turns out to be equal to zero. The extension of the investigation to
SU(3) will be found elsewhere.  

Since we will calculate $h_{\pi NN}^1$ in the process of
$n\pi^+\rightarrow p$, we can rewrite the LO and NLO Lagrangians in 
SU(2) as follows: 
\begin{eqnarray}
{\cal L}^\pi_{\rm LO} &=& i\frac{\sqrt{2}}{f_\pi} \left\{
(- \tilde{\alpha}_{11} + 2 \tilde{\gamma}_{22} )
\left[V^3_{\mu}V^{+ \mu}
+ A^3_{\mu}A^{+ \mu}
\right] + 2(3 \tilde{\gamma}_{21} - \tilde{\gamma}_{22})
V^0_\mu V^{+ \mu}
\right. \nonumber \\
& & \left.+\, 2(3\tilde{\gamma}_{12} - \tilde{\gamma}_{22})
A^0_\mu A^{+ \mu}
+ 2\tilde{\gamma}_{22} \left[
A^0_\mu A^{+ \mu}
+ V^0_\mu V^{+ \mu}
\right]
\right\}\pi^- +\bigg({\bm O}^{0,3} \leftrightarrow {\bm O}^+ \bigg)
\\
{\cal L}_{\rm NLO}^\pi &=& i \frac{\sqrt{2}}{N_c \, f_\pi}
\bigg[
- (\tilde{\alpha}_{11}+2\tilde{\beta}_{11}) \bm{\Lambda}_3^+
+ 3 \left(\frac{4{\cal I}_1{\cal I}_3}{{\cal I}_2^2} + 1 \right)
(\tilde{\gamma}_{12}+2\tilde{\rho}_{12}) \bm{\Lambda}_0^+
\nonumber\\ &&
+\, 3 \left(\frac{4{\cal I}_1{\cal I}_3}{{\cal I}_2^2} - 1 \right)
(\tilde{\gamma}_{21}+2\tilde{\rho}_{21}) \bm{\Lambda}_0^+
+ 2 (\tilde{\gamma}_{22}+2\tilde{\rho}_{22})
(\bm{\Lambda}_3^+ +\bm{\Lambda}_0^+ )\bigg]\pi^-\ ,
\end{eqnarray}
where we have used the identity
\begin{equation}
(\bm{O}\times\bm{\pi})^3= \sqrt{2}i(\pi^-\bm{O}^+-\pi^+\bm{O}^-)
\end{equation}
with the definitions
$\bm{O}^\pm=\frac{1}{2}(\bm{O}^1\pm i\bm{O}^2)$ and
$\pi^\pm = \frac{1}{\sqrt{2}} (\pi_1 \pm i \pi_2)$.
The eighth component of the Gell-Mann matrices becomes the unity
matrix with factor $1/\sqrt{3}$ in going from SU(3) to SU(2).
The PV $\pi NN$ coupling constant, $h_{\pi NN}^1$, can be directly read
from the matrix element
\begin{equation}
h_{\pi NN}^1=i\langle p\uparrow|{\cal L}_{\rm PV}^\pi|n\uparrow,\pi^+\rangle\ ,
\end{equation}
where
\begin{equation}
{\cal L}_{\rm PV}^\pi=-h_{\pi NN}^1{\bar\Psi}_Ni({\bm \pi}^- \tau^+
-{\bm \pi}^+\tau^-)\Psi_N.
\end{equation}

Let us first compute $h_{\pi NN}^1$ with the LO Lagrangian.
Note that the iso-scalar current vanishes identically in the present
model. By using the results in the previous Section, we can see that
the temporal component can contribute to $h_{\pi NN}^1$
because of the orthogonality of $D^{ab}$. This produces the following
expression:
\begin{eqnarray}
V_0^3V^+_0&=&\frac{4 f_\pi^4}{15\sqrt{2}}\sin^4P(r) \cr
&\times & \bigg[6D^{3i}\Omega^i(D^{1j}+iD^{2j})\Omega^j
+D^{3i}\Omega^j(D^{1i}+iD^{2i})\Omega^j
+D^{3i}\Omega^j(D^{1j}+iD^{2j})\Omega^i\bigg]\ ,
\nonumber\\
A_0^3A^+_0&=&\frac{4 f_\pi^4}{3\sqrt{2}}\sin^2 P(r)\cos^2 P(r)
\bigg[D^{3i}\Omega^j(D^{1i}+iD^{2i})\Omega^j
-D^{3i}\Omega^j(D^{1j}+iD^{2j})\Omega^i\bigg]\ .
\label{eq:a0a3}
\end{eqnarray}
Because of the zero-mode quantization, the angular velocity is
expressed in terms of the spin operator $S^i$ $\Omega^i=S^i/I$,
where $I$ is the moment of inertia of the soliton.
The spin operator and Wigner $D$ function satisfy the commutation
relation $[S^i, D^{aj}]=i\epsilon^{ijk}D^{ak}$. Then, the matrix
elements of Eq.(\ref{eq:a0a3}) are written as 
\begin{eqnarray}
\langle p\uparrow|V_0^3V^+_0|n\uparrow\rangle&=&
\frac{f_\pi^4}{15\sqrt{2}} \, \frac{5}{2I^2} \,
\sin^4 P(r)=-\langle p\uparrow|V^+_0V_0^3|n\uparrow\rangle ,
\nonumber\\
\langle p\uparrow|A_0^3A^+_0|n\uparrow\rangle&=&
-\frac{f_\pi^4}{3\sqrt{2}}\, \frac{3}{2I^2} \,
\sin^2 P(r) \cos^2 P(r)=-\langle p\uparrow|A^+_0A_0^3|n\uparrow\rangle.
\end{eqnarray}
Since the LO Lagrangian is symmetric under the exchange of
the indices $3$ and $+$, it turns out that 
\begin{equation}
h_{\pi NN}^1({\rm LO})=0\ \ .
\label{eq:h1pilo}
\end{equation}
This null result of the LO $h_{\pi NN}^1$ was also obtained in the
minimal Skyrme model~\cite{Shmatikov:1989mf}. 

The NLO Lagrangian has a rather complicated structure, so that it is
convenient to analyze first $\bm{\Lambda}_{0,3}^i$. Introducing
$r^i_\mu$ and $l^i_\mu$ as 
\begin{equation}
R_\mu=-\tau^i r^i_\mu, \ \ \ L_\mu=-\tau^i l^i_\mu \, ,
\end{equation}
we rewrite the expressions for $\bm{\Lambda}^i_j$ as
\begin{eqnarray}
\bm{\Lambda}^i_a &=& \frac{f^4_\pi}{4}
\mbox{Tr}[(R_\mu\tau_aR^\mu+L_\mu\tau_aL^\mu) \tau^i ]
= \frac{f^4_\pi}{4} (r^m_\mu r^{n \mu}+l^m_\mu l^{n \mu})
{\rm Tr}(\tau^m\tau_a\tau^n\tau^i)\cr
&=& \frac{f^4_\pi}{2} \bigg(r_{a\mu} r^{i \mu}-\delta^i_ar^m_\mu r^{m \mu}
+ r^{i \mu}r_{a\mu} +(r\leftrightarrow l)\bigg)\,, \mbox{ for }
a\neq 0\, ,
\\
\bm{\Lambda}^i_0 &=&
\frac{f^4_\pi}{12} \mbox{Tr} [(R_\mu R^\mu+L_\mu L^\mu) \tau^i]
= \frac{f^4_\pi}{12} (r^m_\mu r^{n \mu}+l^m_\mu l^{n \mu})
{\rm Tr}(\tau^m\tau^n\tau^i)
\nonumber\\ &=& i \,
\frac{f^4_\pi}{6} \epsilon^{mni}(r^m_\mu r^{n \mu}+l^m_\mu l^{n
  \mu})\,.
\end{eqnarray}
Here, index $a$ runs over $a=1,\, 2,\, 3$. Then, $r^i_\mu$ and
$l^i_\mu$ become
\begin{eqnarray}
r^a_0&=&-D^{ab}(-\sin P(r) \cos P(r)\epsilon^{blm}\Omega^l\hat{r}^m
+\sin^2 P(r) \delta_T^{bl}\Omega^l)\ ,
\\
r^a_i&=&-D^{ab}\bigg(\hat{r}^i\hat{r}^b\partial_r P(r)
+\delta_T^{bi}\frac{\sin2 P(r)}{2r}+\frac{\sin^2 P(r)}{r}
\epsilon^{ikb}\hat{r}^k\bigg)\
,
\\
l^a_0&=&-D^{ab}(\sin P(r) \cos P(r) \epsilon^{blm}\Omega^l\hat{r}^m
 + \sin^2 P(r)\delta_T^{bl}\Omega^l)\ ,
\\
l^a_i&=&-D^{ab}\bigg(-\hat{r}^i\hat{r}^b\partial_r P(r)
-\delta_T^{bi}\frac{\sin2 P(r)}{2r}+\frac{\sin^2
  P(r)}{r}\epsilon^{ikb}\hat{r}^k\bigg)\,,
\end{eqnarray}
where the transverse Kronecker delta is expressed as
$\delta_T^{ab}=\delta^{ab}-\hat{r}^a\hat{r}^b$. Putting these
results together, we arrive at the expressions for ${\bm \Lambda}_3^+$
and ${\bm \Lambda}_0^+$:
\begin{eqnarray}
{\bm \Lambda}_3^+ &=&\frac{f^4_\pi}{4} 
\mbox{Tr}[(R_\mu \tau_3 R^\mu+L_\mu \tau_3 L^\mu) \tau^+ ]
=\frac{f_\pi^4}{4}
\bigg((r^1_\mu+ir^2_\mu)r^{3\mu}+r_\mu^3(r^{1\mu}+ir^{2\mu})+(r
\rightarrow l)\bigg), 
\nonumber\\
{\bm \Lambda}_0^+ &=&\frac{f^4_\pi}{4} 
\mbox{Tr} [(R_\mu\lambda_0 R^\mu+L_\mu\lambda_0 L^\mu) \tau^+ ]
=\frac{f_\pi^4}{12}
\bigg((r^1_\mu+ir^2_\mu)r^{3\mu}-r_\mu^3(r^{1\mu}+ir^{2\mu})+(r
\rightarrow l)\bigg)\ . 
\end{eqnarray}
Since 
\begin{eqnarray}
\int d^3x\, \langle p \uparrow
|r_\mu^3(r^{1\mu}+ir^{2\mu})|n\uparrow\rangle 
&=&-\int d^3x\, \langle p \uparrow
|(r^{1\mu}+ir^{2\mu})r_\mu^3|n\uparrow\rangle 
\nonumber\\
&=&\frac{2\pi}{3I^2}\int dr\,r^2 \sin^2P(r) (\sin^2P(r) -3\cos^2P(r)) , 
\end{eqnarray}
one can easily see that only $\bm{\Lambda}_0^+$ contributes to $h_{\pi
  NN}^1$.
As a result, $h^1_{\pi NN}$ from the NLO Lagrangian
turns out to be
\begin{eqnarray}
h_{\pi NN}^1({\rm NLO}) &=&
\frac{8\sqrt{2} \pi }{3 f_\pi I^2} \,\,
\left({\cal N}_9 + \frac{2}{3} {\cal N}_{10}\right)
\int dr\,r^2\,\sin^2P(r)(\sin^2P(r)-3\cos^2P(r))\, ,
\label{eq:h1pinlo}
\end{eqnarray}
where the LECs ${\cal N}_9$ and ${\cal N}_{10}$ are
given as~\cite{Lee:2004tr}  
\begin{eqnarray}
{\cal N}_9 &=& 4 N_c \left[4 {\cal I}_1 {\cal I}_3 
(\tilde{\gamma}_{12}+ + \tilde{\gamma}_{21}
+2\tilde{\rho}_{12}+2\tilde{\rho}_{21})
+ {\cal I}_2^2
(\tilde{\gamma}_{12} - \tilde{\gamma}_{21}
+2\tilde{\rho}_{12}-2\tilde{\rho}_{21})\right] \cr
&=& 4\langle \overline{\psi}\psi\rangle_M \mathcal{I}_3
\left(\tilde{\gamma}_{12} + \tilde{\gamma}_{21} + 2\tilde{\rho}_{12} +
    2\tilde{\rho}_{21}\right) +
  \frac{f_\pi^4}{4N_c}\left(\tilde{\gamma}_{12} - \tilde{\gamma}_{21} +
    2\tilde{\rho}_{12} -  2\tilde{\rho}_{21}\right), 
\label{eq:n9}\\ 
{\cal N}_{10} &=& 4 N_c\, {\cal I}_2^2 \,
(\tilde{\gamma}_{22}+2\tilde{\rho}_{22}) \;=\; \frac{f_\pi^4}{4N_c}
\left (\tilde{\gamma}_{22} + \tilde{\rho}_{22}\right). 
\end{eqnarray}
As we will discuss later, the LECs $\mathcal{N}_9$ and
$\mathcal{N}_{10}$ are essential to describe the PV $\pi NN$ coupling
constant. 

\section{Results and discussion\label{sec:res}}
We are now in a position to calculate Eq.(\ref{eq:h1pinlo})
numerically. In doing so, we make use of the momentume dependent quark
mass derived from the instanton vacuum~\cite{Diakonov:1985eg} and the
corresponding results of the LECs obtained in
Ref.~\cite{Lee:2004tr}. The value of $M_0=M(k=0)$ is taken to be $350$
MeV as in Ref.~\cite{Lee:2004tr}, which was fixed by the saddle-point
equation from the instanton vacuum~\cite{Diakonov:1985eg}.
Moreover, we employ three different types of
the solitonic profile function to examine the dependence of $h_{\pi
  NN}^1$ on them. The first one is the arctangent profile
function $P(r)$~\cite{diako98} 
\begin{equation}
P(r) = 2 \arctan \left(\frac{r_0}{r}\right)^2,
\label{eq:arctan}
\end{equation}
where $r_0$ is given by $r_0 = \sqrt{\frac{3 g_A}{16 \pi f^2_\pi}}$. 
Employing $g_A = 1.26$ and $f_\pi = $ 93 MeV, we obtain $r_0 = 0.582$
fm. We use a physical profile function as a second one, which
associates with the proper pion tail of the nucleon 
\begin{eqnarray}
P(r) = \left\{
\begin{array}{ll}
2 \arctan\left(\frac{r_0}{r} \right)^2 & (r \leq r_x) \\
A\, {\rm e}^{- m_\pi r} (1 + m_\pi r)/r^2 & (r > r_x),
\end{array}
\right.
\label{eq:hybrid}
\end{eqnarray}
where $m_\pi$ denotes the pion mass and $A = 2 r^2_0$. 
$r_x$ is determined by the intersection of the arctangent function
($r\leq r_x$) and pion tail ($r > r_x$). If one takes the limit $m_\pi
\rightarrow 0$ for the pion tail, the physical profile function
becomes identical with the arctangent one at large $r$. With the
physical pion mass considered, we have $r_x = 0.749$ fm. The final one
is the linear profile function initially proposed by
Skyrme~\cite{skyrme61-1}    
\begin{eqnarray}
P(r) = \left\{
\begin{array}{ll}
\pi (1 - u / \lambda) & (u \leq \lambda) \\
0 & (u > \lambda)
\end{array}
\right.
\label{eq:stick}
\end{eqnarray}
where $u \equiv 2 e f_\pi r$ with $e = 4.84$ and $\lambda = 3.342$. 

\begin{figure}[h]
\centerline{
\includegraphics[scale=0.8]{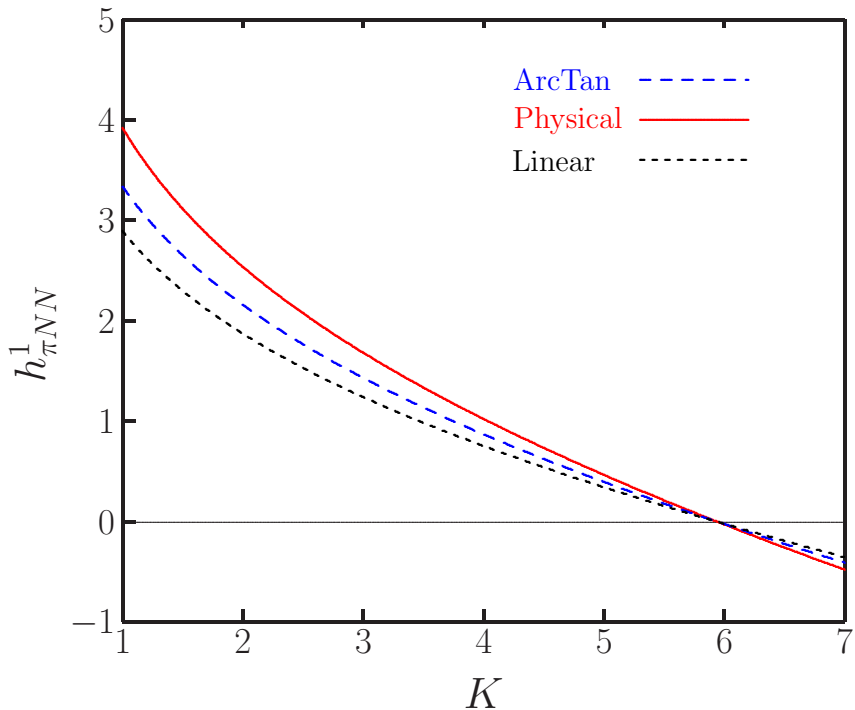}
}
\caption{(Color online) PV $\pi NN$ coupling constant $h^1_{\pi NN}$
  as a function of the Wilson coefficient $K$ in units of $10^{-8}$. The
  solid curve draws the result with the physical profile function, the
dashed one depicts that with the arctangent profile, and the
short-dashed one does that with the linear one. }  
\label{fig:1}
\end{figure}

Using these three profile functions, we can immediately compute 
the PV $\pi NN$ coupling constant $h^1_{\pi NN}$. Figure~\ref{fig:1} 
draws the results of $h^1_{\pi NN}$ as a function of the Wilson coefficient
$K$. 
The solid curve depicts that with the physical profile function,
whereas the dashed and short-dashed ones correspond to those with the
arctangent and linear profile functions respectively. One can regard the
difference between the results with the physical profile function and
those with the arctangent one as effects of the finite pion mass,
which contribute to $h_{\pi NN}^1$ approximately by $10\,\%$. Moreover, the
type of the profile function does not change much the general features
of $h_{\pi NN}^1$, though we preferably take the results with the physical
one as our final values.  

\begin{figure}[ht]
\centerline{\includegraphics[scale=0.8]{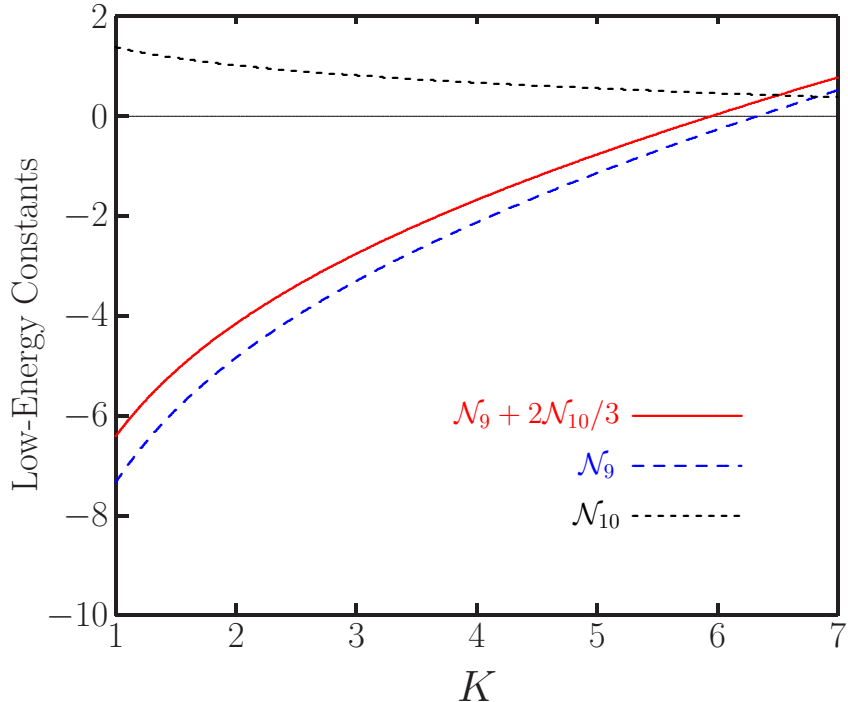}}
\caption{(Color online) Low-energy constants $\mathcal{N}_9$,
$\mathcal{N}_{10}$, and $\mathcal{N}_9 + 2\mathcal{N}_{10}/3$ as 
functions of the Wilson coefficient $K$ in units of
$10^{-11}\,\mathrm{GeV}^2$. The solid curve draws the result of
$\mathcal{N}_9 + 2\mathcal{N}_{10}/3$, 
dashed one depicts that of $\mathcal{N}_9$, and 
the short-dashed one does that of $\mathcal{N}_{10}$.}
\label{fig:2}
\end{figure}

We find out from Fig.~\ref{fig:1} that $h_{\pi NN}^1$ is rather sensitive
to the Wilson coefficient $K$ and it decreases monotonically, as $K$
increases.  We notice that its sign is even changed around
$K=6$. This can be easily understood. The LECs
$\mathcal{N}_9$ and $\mathcal{N}_{10}$ in Eq.~(\ref{eq:h1pinlo}) play
essential roles in determining the $K$ dependence of
$h_{\pi NN}^1$. Figure~\ref{fig:2} draws the results of the LECs
$\mathcal{N}_9$ and $\mathcal{N}_{10}$ as functions of 
$K$. While $\mathcal{N}_9$ depends rather strongly on $K$,
$\mathcal{N}_{10}$ does mildly on $K$. Moreover, $\mathcal{N}_9$ is
dominant over $\mathcal{N}_{10}$, so that the PV $\pi NN$ coupling 
constant is mainly governed by $\mathcal{N}_9$. 
Since $\mathcal{N}_9$ is the main contribution to $h_{\pi NN}^1$, we want
to examine it in detail. We can easily see that
the first term of Eq.(\ref{eq:n9}) containing the quark condensate is
much larger than the second one. Moreover, since $\tilde{\gamma}_{12}$
and $\tilde{\rho}_{21}$ are much smaller than the other two
coefficients $\tilde{\gamma}_{21}$ and $\tilde{\rho}_{12}$, we can
neglect them in $\mathcal{N}_9$. Then, $\mathcal{N}_9$ can be
expressed as 
\begin{equation}
\mathcal{N}_9 \;\approx\; 4\langle \overline{\psi}\psi\rangle_M
\mathcal{I}_3 \left(\tilde{\gamma}_{21} + 2\tilde{\rho}_{12} 
\right). 
\end{equation}
As shown in Eq.(\ref{Eq:Heff}), the coefficient
$\tilde{\gamma}_{21}$ comes from the original effective weak
Hamiltonian at the mass scale of the $W$ boson
$\mu=M_W=80.4\,\mathrm{GeV}$ corresponding to 
$K=1$.  In this case, only $\tilde{\gamma}_{21}$ survives in
$\mathcal{N}_9$.  However, when we start to scale the Hamiltonian
down to $\mu\approx 1\,\mathrm{GeV}$ that corresponds to $K\approx  
4$, the gluonic renormalization arising from gluon exchange parallel
to $Z$-boson exchange is turned on. As a result, the
$2\tilde{\rho}_{12}$ term becomes as large as a 
half of the $\tilde{\gamma}_{21}$
one~\cite{Desplanques:1979hn,Lee:2004tr} at this scale. If one goes 
further down to the scale at which $K\approx 6$~\footnote{There is a
caveat in scaling further down below 1 GeV, because the matching
problem becomes non-trivial below the charm quark mass. Furthermore,
we still do not know how to incorporate all possible nonperturbative
effects consistently below 1 GeV.},  
the correction of $\tilde{\rho}_{12}$ cancels out the contribution
of $\tilde{\gamma}_{21}$, so that $h_{\pi NN}^1$ almost vanishes, as
already shown in Figs.~\ref{fig:1} and \ref{fig:2}. 
This cancellation implies that the effects of the gluon
renormalization leads to the suppression of the PV $\pi NN$ coupling
constant in the present approach of the SU(2) $\chi$QSM.  

Since Ref.~\cite{Lee:2004tr} derived the effective weak chiral
Lagrangian based on the $\Delta S=0$ effective weak Hamiltonian, it is
plausible to take the value $K=4$ for $h_{\pi NN}^1$, 
which corresponds to the renormalization scale of the charm quark mass
$\mu \approx 1$ GeV as done for the $\Delta S=1$
case~\cite{Buchalla:1995vs}.  Thus, we obtain $h_{\pi NN}^1\approx1\times
10^{-8}$ for $K=4$.  However, if one
neglects all the renormalization 
effects, i.e. if one takes $K=1$, we have $h_{\pi NN}^1\approx 4 \times
10^{-8}$, which is similar to that of the SU(2) Skyrme model with
vector mesons ($h_{\pi NN}^1=(2-3)\times 10^{-8}$)~\cite{Kaiser:1989fd} in
which the effective weak Hamiltonian at $\mu=M_W$ or with $K=1$ was
used. Note that the present result is almost 40 times smaller than the
``best'' value of DDH ($h_{\pi NN}^1=4.5\times
10^{-7}$)~\cite{Desplanques:1979hn}. 

\begin{figure}[ht]
\centerline{\includegraphics[scale=0.8]{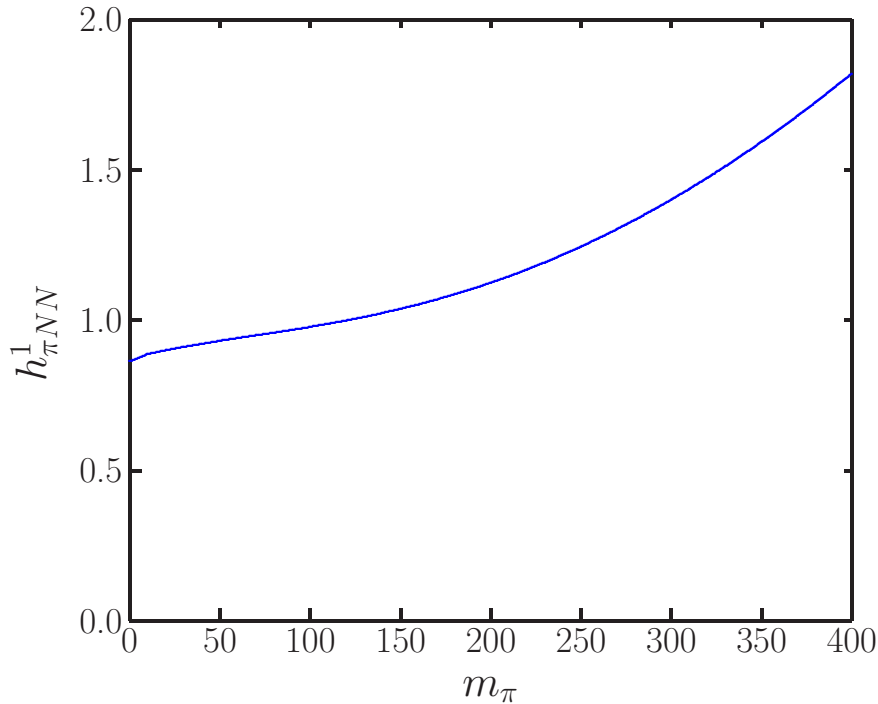}}
\caption{(Color online) PV $\pi NN$ coupling constant as a function of
  the pion mass $m_\pi$ in units of $10^{-8}$. The Wilson coefficient
$K=4$ is used, which 
  corresponds to $\mu \approx 1$ GeV.}
\label{fig:3}
\end{figure}
Very recently, Ref.~\cite{Wasem:2011tp} has reported the first result
of lattice QCD: $h_{\pi  NN}^1=(1.099\pm0.505_{-0.064}^{+0.058})\times
10^{-7}$ with the pion mass $m_\pi = 389$ MeV. Thus, it is 
interesting to compare the present result with the lattice one. In
order to do that, we need to examine the dependence of $h_{\pi NN}^1$
on the pion mass $m_\pi$. Figure~\ref{fig:3} depicts the PV $\pi NN$
coupling constant as a function of $m_\pi$. We employ here the
physical profile function with $K=4$. Interestingly, $h_{\pi NN}^1$
starts to increase, as $m_\pi$ does. As a result, $h_{\pi NN}^1$ turns
out to be around $1.8\times 10^{-8}$ for $m_\pi=400$ MeV. Though it is
still around five times less than that of the lattice calculation, we
can infer from Fig.~\ref{fig:3} 
that lattice results with the physical pion mass might be quite
smaller than that of Ref.~\cite{Wasem:2011tp}. Moreover, if one takes
$K=1$, $h_{\pi NN}^1$ with $m_\pi=389$ MeV would become $h_{\pi
  NN}^1\approx 6.77\times 10^{-8}$ that is comparable to the lattice
one, though the value $K=1$ does not seem tenable for $h_{\pi NN}^1$
as discussed previously. 
However, one has to keep in mind that Ref.~\cite{Wasem:2011tp} has not 
performed the calculation of the full matrix element, since the
quark-loop diagrams were omitted because of technical difficulties.   
A quantitative comparison with full lattice calculations is still
being awaited.  

\section{Summary and outlook}
We have investigated the parity-violating pion-nucleon coupling
constant $h_{\pi NN}^1$ within the framework of the chiral quark-soliton
model with the gradient expansion used. Starting from the $\Delta S=0$
effective weak chiral Lagrangian derived in the same framework, we
have calculated the parity-violating $\pi NN$ coupling constant.
It was found that it vanished at the leading order in the large
$N_c$, i.e. $h_{\pi NN}^1(\mathrm{LO}) = 0$, which is of order
$\mathcal{O}(N_c^{-1/2})$, but it was finite to the next-to-leading
order, i.e. $\mathcal{O}(N_c^{-3/2})$. 

Employing three different profile functions, that is, the arctangent,
physical, and linear ones, we calculated the parity-violating $\pi NN$
coupling constant to the next to the leading oder. It turns out that
the values of $h_{\pi NN}^1$ depend sensitively on the values of the Wilson
coefficient $K$ and vanishes around $K\approx 6$. The reason can be
found in the fact that the contribution of the gluonic renormalization
constant $\tilde{\rho}_{12}$ cancels out that of the
$\tilde{\gamma}_{21}$, which is the leading one. It indicates that the
perturbative gluonic contribution supresses the parity-violating $\pi
NN$ coupling constant.   
Taking the scale of the charm quark mass, i.e. $\mu\approx 1$ GeV, 
we found $h_{\pi NN}^1\approx 1\times 10^{-8}$, which is almost 40 times
smaller than the ``best value'' of Ref.~\cite{Desplanques:1979hn}.   
If the $\mu=M_W$ is selected, the value of $h_{\pi NN}^1$ turns out to be
similar to that from the Skyrme model with vector
mesons~\cite{Kaiser:1989fd}. 
We also compared the present result with that of the lattice
calculation. Thus, we examined the dependence of the 
parity-violating $\pi NN$ coupling constant on the pion mass and found
that $h_{\pi NN}^1$ increased as $m_\pi$ did. If one uses $m_\pi=400$
MeV, the result turns out to be almost two times larger than that with
the physical value $m_\pi=140$ MeV but is still about five times
smaller than the lattice one. 
However, we want to emphasize that
neither the present result nor the lattice one is the final
one. 

In order to understand the parity-violating $\pi NN$ coupling
constant $h_{\pi NN}^1$ more completely and quantitatively, we have to
consider the following important physics:
Since we have considered the SU(2) case in the present work, the
effects of strangeness were left out. As already mentioned in
Section~\ref{sec:lag} , however, the strange quark operators may play
a certain role in describing the parity-violating $\pi NN$
coupling constant. Extending from SU(2) to SU(3) is lengthy but
straightforward in the present framework. Starting from
Eqs.(\ref{eq:pilag2}, \ref{eq:NLO}), we employ the quantization with
the embedding~(\ref{eq:embed}). In particular, the singlet current is  
distinguishable from the octet one in SU(3), so that this would make
difference in predicting the parity-violating $\pi NN$ coupling
constant.  Moreover, the fourth and fifth flavor components of the
vector currents enter the next-to-leading order Lagrangian, this would
also contribute to $h_{\pi NN}^1$.   

As was seen in Section~\ref{sec:res}, the gluon renormalization
plays a role of suppressing the parity-violating $\pi NN$ coupling
constant. However,the effective weak Hamiltonian at two-loop order
was derived very recently in Ref.~\cite{Tiburzi:2012hx}, where the
QCD penguin diagrams were also considered. This Hamiltonian is more
complete than that from Ref.~\cite{Desplanques:1979hn}. Thus, it is
of great significance to investigate the $\Delta S=0$ effective weak
chiral Lagrangian based on this effective weak Hamiltonian. It is also
of great interest to see whether these penguin diagrams enhance the
$h_{\pi NN}^1$ or not, since they give part of answers of explaining
the $\Delta I=1/2$ rule in nonleptonic
decays~\cite{Shifman:1975tn,Buchalla:1995vs}.  The corresponding
investigations are under way.

\begin{acknowledgments}
H.Ch.K is grateful to P. Schweitzer for his hospitality during his visit
to Department of Physics, University of Connecticut, where part of the
present work was carried out. The work of H.Ch.K. was supported by
Inha University Research Grant. This work was also
supported by the research grant of the Chungbuk National University in
2011 (HJL).  
\end{acknowledgments}

\end{document}